\documentclass[12pt]{article}

\usepackage{graphics,graphicx,color,float, hyperref}
\usepackage{bm}
\usepackage{mathtools}
\usepackage{amsmath}
\usepackage{amssymb}
\usepackage{amsfonts}
\usepackage{amsthm}
\usepackage {times}
\usepackage{layout}
\usepackage{setspace}
\usepackage{geometry}
\usepackage{tikz-cd}

\usepackage[margin=1cm]{caption}

\newcommand{\beq}{\begin{equation}}
\newcommand{\enq}{\end{equation}}
\newcommand{\beqst}{\begin{equation*}}
\newcommand{\enqst}{\end{equation*}}
\newcommand{\beqar}{\begin{eqnarray}}
\newcommand{\enqar}{\end{eqnarray}}
\newcommand{\beqarst}{\begin{eqnarray*}}
\newcommand{\enqarst}{\end{eqnarray*}}
 
\newcommand*{\cL}{\mathcal{L}}
\newcommand*{\cC}{\mathcal{C}}

\newcommand*{\cH}{\mathcal{H}}

\newcommand*{\cF}{\mathcal{F}}

\newcommand*{\cD}{\mathcal{D}}

\newcommand*{\cN}{\mathcal{N}}

\newcommand*{\cE}{\mathcal{E}}

\newcommand{\cP}{\mathcal{P}}

\newcommand{\bR}{\mathbb{R}}
\newcommand{\bX}{\textbf{X}}
\newcommand{\bC}{\textbf{C}}

\newcommand{\bbN}{\mathbb{N}}

\newcommand{\Tr}{\mathrm{Tr}}
\newcommand{\eps}{\varepsilon}

\newcommand{\suppress}[1]{}

\newcommand{\F}{\mathrm{F}}
\newcommand {\br} [1] {\ensuremath{ \left( #1 \right) }}

\newcommand {\minusspace} {\: \! \!}
\newcommand {\smallspace} {\: \!}
\newcommand {\fn} [2] {\ensuremath{ #1 \minusspace \br{ #2 } }}
\newcommand {\ball} [2] {\fn{\mathcal{B}^{#1}}{#2}}
\newcommand {\ent} [1] {\mathrm{S}(#1)}
\newcommand {\cent} [2] {\mathrm{S}\br{#1 | #2}}
\newcommand {\relent} [2] {\fn{\mathrm{D}}{#1 \middle\| #2}}
\newcommand {\mutinf} [2] {\fn{\mathrm{I}}{#1 \smallspace : \smallspace #2}}
\newcommand {\condmutinf} [3] {\mutinf{#1}{#2 \smallspace \middle\vert \smallspace #3}}
\newcommand {\id}{\mathrm{I}}

\newcommand{\ketbra}[1]{|#1\rangle\langle#1|}
\newcommand{\bra}[1]{\langle #1|}
\newcommand{\ket}[1]{|#1 \rangle}

\mathchardef\mhyphen="2D

\makeatletter
\newcommand*{\rom}[1]{\expandafter\@slowromancap\romannumeral #1@}
\makeatother

\mathchardef\mhyphen="2D

\newtheorem{theorem}{Theorem}
\newtheorem{lemma}{Lemma}

\newtheorem{fact}{Fact}

\newtheorem{definition}{Definition}
\newtheorem{claim}{Claim}

\newcommand{\change}[1]{#1}
\newcommand{\newchange}[1]{#1}

\begin {document}
\title{Incompressibility of classical distributions}
\author{
Anurag Anshu\thanks{Department of EECS, University of California, Berkeley and Simons Institute for the Theory of Computing, Berkeley} \qquad
Debbie Leung\thanks{IQC and C\&O, University of Waterloo, Perimeter Institute} \qquad
Dave Touchette\thanks{IQ and CS, Universite de Sherbrooke, IQC and C\&O, University of Waterloo, Perimeter Institute} \qquad
}

\hypersetup{pageanchor=false} 
\date{}
\maketitle

\vspace*{-2ex}

\abstract{In \emph{blind} compression of quantum states, a sender Alice is given a specimen of a quantum state $\rho$ drawn from a known ensemble (but without knowing what $\rho$ is), and she transmits sufficient quantum data to a receiver Bob so that he can decode a near perfect specimen of $\rho$.  For many such states drawn iid from the ensemble, the asymptotically achievable rate is the number of qubits required to be transmitted per state.  The Holevo information is a lower bound for the achievable rate, and is attained for pure state ensembles, or in the related scenario of entanglement-assisted \emph{visible} compression of mixed states wherein Alice knows what state is drawn.

In this paper, we prove a \change{general and robust} lower bound on the achievable rate for ensembles of classical states, which holds even in the least demanding setting when Alice and Bob share free entanglement and a constant per-copy error is allowed.  We apply the bound to a \emph{specific} ensemble of only two states and prove a near-maximal separation \change{(saturating the dimension bound in leading order)} between the best achievable rate and the Holevo information for constant error.  \change{This also implies that the ensemble is incompressible -- compression does not reduce the communication cost by much.}  Since the states are \emph{classical}, the observed incompressibility is not fundamentally quantum mechanical.  We lower bound the difference between the achievable rate and the Holevo information in terms of quantitative limitations to clone the specimen or to distinguish the two classical states.} 

\section{Introduction}

\subsection{Blind quantum data compression and related scenarios}

A central goal of information theory is to capture the ultimate rate of transformation of resources.  For example, we may want to minimize the communication cost of a task, which is an optimization problem over a potentially unbounded number of possible communication protocols.  In some special cases, the best communication cost is given by a simple enough information theoretic quantity that can be computed.  For example, this has been achieved in Shannon's source coding theorem (data compression) and noisy channel coding \cite{Shannon} and some network analogues \cite{GamalK12}.  Quantum information theory shares the same goal, and similar understanding has been achieved in quantum noisy coding theorem, albeit with regularization issues in many scenarios.

This paper focuses on the problem of quantum data compression, which can be stated as follows.
Fix an ensemble $\{p(x), \rho^x_C\}$ of quantum states $\rho^x_C$ on a register $C$, and define the associated state
\begin{equation}
\rho_{XC} = \sum_xp(x)\ketbra{x}_X\otimes\rho^x_C \,.
\label{rhoxc}
\end{equation}
In the aforementioned ensemble, each state $\rho^x_C$ is labeled by a classical index $x$ recorded in the register $X$ and occurring with probability $p(x)$.  Suppose a Referee prepares $n$ copies of the above state:
\begin{equation}
\sum_{x_1 x_2 \cdots x_n} p(x_1)p(x_2) \cdots p(x_n) \;
\ketbra{x_1}_{X_1} \! \otimes
\ketbra{x_2}_{X_2} \! \otimes
\cdots \otimes 
\ketbra{x_n}_{X_n} \otimes
\rho^{x_1}_{C_1} \otimes \rho^{x_2}_{C_2} \otimes \cdots \otimes \rho^{x_n}_{C_n},
\label{rhoxcn}
\end{equation}
where each $C_i\equiv C$.  The Referee transmits $C_1 \cdots C_n$ to
Alice.  Alice is allowed to send some quantum data to Bob.  Bob
decodes and his final output registers are $C_1' \cdots C_n'$.  The
goal is that the final state of the Referee and Bob should be
\emph{close} to the state in (\ref{rhoxcn}), while minimizing the amount
of data sent.  A rate $r$ is achievable for the compression if there
is a family of protocols labeled by $n$ in which Alice sends $nr$
qubits.  There are many related but inequivalent scenarios for quantum
data compression.
\begin{itemize}
\item
In the \emph{blind} scenario (as described above), Alice
does not have access to the registers $X_1 \cdots X_n$. She has a
specimen of the states $\rho^{x_1} \otimes \cdots \otimes \rho^{x_n}$
in $C_1 \cdots C_n$, but she does not known what they are in general.
In contrast, in the \emph{visible} scenario, the Referee gives a copy
of $X_1 \cdots X_n$ to Alice so she knows $x_1 \cdots x_n$ (and in
this case it is unnecessary to give her $C_1 \cdots C_n$).
\item
In the unassisted model, Alice and Bob do not share any
correlations.  In other scenarios, they may share classical
randomness.  In the entanglement-assisted scenario, they may
share any entangled state of their choice.
Note that with entanglement assistance, sending quantum or classical data are equivalent
due to teleportation \cite{Teleportation93} and superdense coding \cite{BennettW92}.
The rate in qubits is equal to half of the rate in bits.
\item
One has to specify the measure of proximity between the initial state 
(\ref{rhoxcn}) and the final state held by the Referee and Bob.  
A more stringent definition of error requires that the final state in
$X_1 \cdots X_n C_1' \cdots C_n'$ be close to the original state in
(\ref{rhoxcn}), in trace distance or in fidelity.  The error in this
case is called ``global.''  
A more relaxed definition of error requires that for each $i$, the
$i$-th output state in $X_i C_i'$ is close to the initial state in
$X_i C_i$.  The error in this case is called ``local.''
For the asymptotic case, the error is typically required to vanish as $n$ increases.  
Alternatively, one can consider the one-shot scenario when $n=1$, 
but this scenario is out of the scope for our paper. We will mention some one-shot results which have asymptotic implications.

\item 
There is no limitation on the states $\rho^x_C$ in the ensemble in the
problem.  There are several special cases of interest.  One
well-studied special case is the ``pure state case'' in which all
$\rho^x_C$ are pure.  Another case concerns ensembles of states that
are commuting, in which case they can be simultaneously diagonalized,
and the $\rho^x_C$ correspond to classical distributions.
\end{itemize}

\noindent We can summarize prior results as follows.  
The unassisted blind scenario for pure state case was formulated 
in \cite{Schumacher95, Josza-Schumacher94, BFJS96}. These pioneering works 
established the quantum analogue of Shannon's source coding theorem when
the ensemble
$\{p(x),\ketbra{\rho^x}_C\}$ consists only of pure states, with 
the best achievable rate shown to be 
$\ent{\rho_C}$ qubits, where $\ent{.}$ is the von-Neumann entropy
\cite{Neumann32} and 
$\rho_C=\sum_xp(x)\ketbra{\rho^x}_C$ is the average state of the ensemble. 
If the states $\ket{\rho^x}_C$
are mutually orthogonal, the problem reduces to Shannon's source
coding problem and Schumacher's protocol recovers Shannon's result
with rate being $\ent{\rho_C}$ bits.

For a general ensemble $\{p(x), \rho^x_C\}$, the Holevo information is
defined as 
$S\!\br{\sum_x \! p(\hspace*{-0.2ex} x \hspace*{-0.2ex})\rho^x_C} - 
  \hspace*{-0.3ex}
  \sum_x \! p( \hspace*{-0.2ex} x \hspace*{-0.2ex}) S 
\hspace*{-0.5ex} \br{\hspace*{-0.2ex}\rho^x_C}$, 
and it is
also the quantum mutual information $\mutinf{X}{C}_{\rho}$ between $X$
and $C$ evaluated on $\rho_{XC}$. It was independently shown by \cite{Horodecki98} and \cite{BCFJS00} that the Holevo
information is a lower bound for the achievable rate in the unassisted
scenarios for both visible and blind compression. 

For unassisted compression of pure states, the above lower
bound on the rate is already attained by the protocol in the blind
setting \cite{Schumacher95, Josza-Schumacher94, BFJS96}. Thus
the visible scenario, with Alice's knowledge of the
state to be compressed, surprisingly does not improve the rate.  Furthermore,
shared randomness does not reduce the best achievable rate.

The situation is more complex for the compression of mixed quantum
states.  The problem was considered as early as in \cite{Josza94} and
formulated and studied in detail in a large body of work
\cite{Horodecki00, Dur, KramerSavari, KoashiI1, KoashiI2, KoashiI3,
  Winter, Soljanin02}.
The rate depends on whether the protocol is visible or blind, what kind
of assistance is available, under local or global error, and
whether the ensemble is classical or
quantum, to be discussed as follows.

Visible compression of classical ensemble is relatively well
understood, given the assistance of shared randomness.  The problem is
equivalent to the simulation of classical channels (associated to the
classical Reverse Shannon theorem). Authors in \cite{BennettSST02} and
\cite{Dur} independently showed that the Holevo information is the
achievable rate in bits under \emph{global} error criteria.  Winter
\cite{Winter} further showed that under the local error criteria,
shared randomness is not needed to achieve the Holevo information. It
was also shown in \cite[Theorem 3]{Winter} that the Holevo information
is a lower bound even for asymptotically non-vanishing global
error. This is a notable feature of visible compression: even a
constant global error (for example $\frac{1}{3}$) requires a rate at
least equal to the Holevo information.
Using rejection sampling, the Holevo information was shown to be
achievable in the asymptotic setting \cite{Jain:2003} and one-shot
with expected communication \cite{HJMR10}.

For the visible compression of quantum ensembles without any assistance, Horodecki \cite{Horodecki00} showed that the qubit rate is given by a quantity defined via extensions of the ensemble. Later, Hayashi \cite{Hayashi06} gave a simpler characterization of the qubit rate in terms of the entanglement of purification \cite{Terhal_2002}.
With entanglement assistance, protocols for the visible compression have several guises.  
The first guise is remote state preparation of entangled states between Alice and Bob, first formalized in \cite{BDDSSTW01} and solved in \cite{BHLSW05} with qubit rate $\frac{1}{2}\mutinf{X}{C}_{\rho}$ (subsequently reproduced from a one-shot approach in \cite{AbeyesingheHGW06}).  
The second is via the rejection sampling method and quantum substate theorem \cite{Jain2002,Jain05} which gives a one-shot protocol with asymptotic rate of $\frac{1}{2}\mutinf{X}{C}_{\rho}$ qubits.
The third guise is via the general scenario of the quantum Reverse Shannon theorem \cite{BennettDHSW14, BertaCR11} which also attains the optimal qubit rate of $\frac{1}{2}\mutinf{X}{C}_{\rho}$.  
The first and third methods are entanglement optimal as well.
\newchange{Reference \cite{Hayashibook17}, Section 10.8, also
addresses visible compression.}

Finally, for blind compression of a mixed ensemble, the difference
between the rate of quantum communication and the Holevo information
was termed ``information defect'' by Horodecki \cite{Horodecki98}.
Both \cite{Horodecki98} and \cite{BCFJS00} provided bounds on the
information defect without resolving whether it could be positive.  In
\cite{Dur}, a classical ensemble was presented with an argument
sketching the positivity of its information defect. But the error
criteria in their argument was not made precise. Kramer and Savari
\cite{KramerSavari} also showed a similar result with an error
criteria based on empirical distribution of the outputs. But this
error criteria does not match either the global or local criteria
discussed above.
In a powerful series of results
\cite{KoashiI1, KoashiI2, KoashiI3}, Koashi and Imoto
characterized the optimal rates of 
quantum and classical 
communication, the amount of entanglement required, and
their tradeoffs in blind compression.  This was done  
by a decomposition of the ensemble of states, now colloquially called
the Koashi-Imoto decomposition. Their result requires that the local
error goes to zero in the asymptotic setting and leads to a 
large information defect. An ensemble witnessing this
separation consists of two equiprobable commuting states
\cite{KoashiI2}, and its blind compression requires classical
communication at the rate of the entropy of the average state. 
\change{This ensemble achieves a near-maximal separation between 
achievable rate of communication and the Holevo information of the ensemble in the regime of vanishing error. 
By `near-maximal', we mean that the leading order term in the
separation saturates the dimension bound.}

We motivate this study by asking if the above rate characterization
holds for non-vanishing error; mimicking the feature of visible
compression mentioned earlier. This is a first step towards chalking
out the ``communication versus error'' profile for blind compression
and understanding its strong converse rate. We observe that the
Koashi-Imoto rate characterization is sensitive to the amount of
error.  We highlight this using an example in Appendix
\ref{koashiIrobust}, where we show that blind compression of any
ensemble of two commuting states $\{\rho^0_C, \rho^1_C\}$ with local
error $\eps$ can be achieved with unassisted rate of $2\log\log
\frac{d}{\eps} + 2\log\frac{1}{\eps}+5$ bits ($d$ is the dimension of
register $C$ which is a constant independent of $n$).  
\change{For $\eps \gg \frac{1}{\sqrt{d}}$, our achievable
rate is substantially less than the lower bound of $\log d$ given by
the Koashi-Imoto rate characterization for a generic pair of commuting
states in the vanishing error regime}.  The general compression rate, as a function of $\eps$,
therefore, remains unresolved.

Is there a coding scheme that can even further reduce the rate
exhibited in the aforementioned example for finite $\eps$? For
instance, could the rate depend on the error as
$\mathcal{O}\left(\log\log\frac{1}{\eps}\right)$, as in
\cite{BDDSSTW01, Jain05, HJMR10} using rejection sampling?  Much of
this paper is devoted to showing the contrary. We provide an example
where the rate of $\mathcal{O}\left(\log\frac{1}{\eps}\right)$ is
optimal for a suitable finite choice of local error. 
For this ensemble,
we show a
large and robust lower bound $(\log d) - 7$ for the rate,
while the Holevo information is less than $1$. \change{By `robust', we mean that the lower bound is a simple function of the error.} Thus compression of  
this ensemble does not reduce communication rate in a significant manner 
relative to sending the whole register $C$.
Note that since our lower bound holds for local
error, it also implies the same lower bound for the global
error. Furthermore, our lower bound applies to entanglement-assisted
protocols.
\change{Since we consider entanglement-assisted protocols, our lower bound also applies to shared randomness assisted protocols, such as in \cite[Section 10.8]{Hayashibook17}.} 

\newchange{Our result may also provide insights to rate distortion theory 
for blind compression of distributions.  Rate-distortion theory has
been studied in the quantum setting, for example, see  
\cite{Barnum00,Datta_2013,Wilde_2013}.  These 
prior works use a less stringent error measure than our current work, but preserve the 
purification of the compressed states.  Therefore, 
their lower bounds do not apply to the current setting. 
}

\subsection{Main result, techniques and consequences}

In this work, we show a near-maximal (for the dimension of the states) separation between the achievable rate of \emph{classical communication} for entanglement-assisted blind mixed state compression and the Holevo information. As mentioned earlier, our separation holds for finite (non-vanishing) local error. We establish this separation in two steps.  

In the first step, we consider entanglement-assisted blind mixed state compression of the $n$-copy state in (\ref{rhoxcn}), for ensembles of \emph{classical} states $\rho^x_C$ that are diagonal.
We obtain a \emph{single-letter} lower bound on the \emph{asymptotic} achievable rate $R$:
\beq
\label{eq:roughlb}
R\geq \min_{\cF:C\rightarrow CC'}\br{\condmutinf{C}{C'}{X}_{\cF(\rho)}}+\mutinf{X}{C}_{\rho} - \eps \log |X| - 1,
\enq
where $\rho$ is defined in (\ref{rhoxc}), the \change{quantum TPCP} map $\cF$ takes $\rho_{XC}$ to $\rho_{XCC'}$ and satisfies the constraints 
\begin{enumerate}
\item $\Tr_{C}\cF(\rho_{XC}) \approx \rho_{XC'}$,  
\item $\Tr_{C'}\cF(\rho_{XC})=\rho_{XC}$, 
\end{enumerate}
and the approximation in the first constraint is given by $\eps$. \change{Since the map $\cF$ acts as identity on the register $X$, we have shortened the notation $\id_X\otimes \cF$ to $\cF$ for convenience.}
Note that despite a similarity in form between (\ref{eq:roughlb}) and Lemma 3.1 in \cite{KramerSavari}, our bound is obtained under the local error condition (unlike the empirical error condition of \cite{KramerSavari}).
The expression $\min_{\cF:C\rightarrow CC'}\br{\condmutinf{C}{C'}{X}_{\cF(\rho)}}$ 
approximates the difference between $R$ and the Holevo information $\mutinf{X}{C}_\rho$ in (\ref{eq:roughlb}), with the following noteworthy features:
\begin{itemize}
\item 
If Alice knows $X$ (visible scenario or distinguishable $\rho^x_C$'s),
the expression vanishes which matches previous known bounds.  
Thus this expression represents Alice's lack of knowledge of the label $X$ of the given state.  Our strategy is to prove a large lower bound for the expression.

\item We can view $C$ as the register containing 
the state held by Alice, and $C'$ as the register $C'$ containing the output state held by Bob.   
The first constraint $\Tr_{C}\cF(\rho_{XC}) \approx \rho_{XC'}$ reflects the correctness of the protocol: the state $\Tr_{C}\cF(\rho_{XC})$ between Bob and the Referee (holding $X$) is close to the desired state $\rho_{XC'}$ (with $C$ replaced by $C'$).
The second constraint $\Tr_{C'}\cF(\rho_{XC})=\rho_{XC}$ comes from classicality of the ensemble, which allows Alice to retain a copy of the classical value in register $C$. \change{Being classical, the system C is not disturbed by the measurement that allows Alice to retain the copy of $C$.}
\item 
In Section \ref{sec:singlelettbdd}, we specialize to equiprobable
ensembles of two states, and convert the expression to two simpler lower
bounds given in (\ref{eq:pinskerlb}).
The first lower bound, (\ref{eq:pinskerlb})(c), is the expected distance
(over $x$) between two joint states shared by Alice and Bob.   
The first joint state is the output of the protocol, and the
second state consists of two independent copies of Alice's input
state, one held by each of Alice and Bob.  Thus, the compression
rate is lower bounded by the \emph{inability} to clone the states in
the ensemble.  
The second lower bound, (\ref{eq:pinskerlb})(f), is the gain in
distinguishability between the states for $x=0$ and $x=1$, if two
copies of the states are available instead of one copy.
We note that these two lower bounds involve \newchange{trace
distances on the joint system held by Alice and Bob, and are  
optimized over} arbitrary joint operations by Alice and Bob.
\newchange{These quantities are also easily evaluated.}
\end{itemize}  

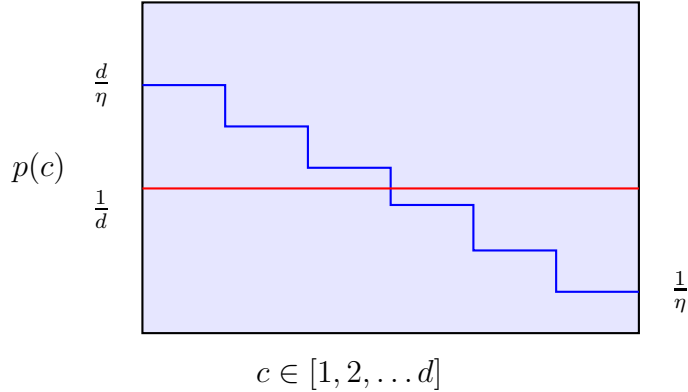
\begin{figure}[ht]
\centering
\begin{tikzpicture}[xscale=0.55,yscale=0.55]
\draw [fill=blue!10!white, thick] (0,0) rectangle (12, 8);
\draw [thick, blue] (0,6) -- (2,6)-- (2,5) -- (4,5) -- (4,4) -- (6,4) -- (6,3.1) -- (8,3.1) -- (8,2) -- (10,2) -- (10,1) -- (12,1);
\draw [thick, red] (0,3.5) -- (12,3.5);
\node at (-2.5, 4) {$p(c)$};
\node at (-1, 3) {$\frac{1}{d}$};
\node at (-1, 6) {$\frac{d}{\eta}$};
\node at (13, 1) {$\frac{1}{\eta}$};
\node at (5, -1) {$c \in [1,2,\ldots d]$};
\end{tikzpicture}
\caption{The two distributions in our example. The red line is the
  uniform distribution and the blue line is the staircase
  distribution.  Here, $d=|C|$ and $\eta=\frac{d(d+1)}{2}$.}
\label{fig:exampledist}
\end{figure}

The lower bounds (\ref{eq:pinskerlb})(c) and (f) are not extensive.  
To obtain a large lower bound on the expression $\min_{\cF:C\rightarrow CC'}\br{\condmutinf{C}{C'}{X}_{\cF(\rho)}}$ limited only by the dimension, we choose an equiprobable ensemble of two states $\rho^0_C$ and $\rho^1_C$, where the former represents the uniform distribution and the latter the `staircase' distribution; see Figure \ref{fig:exampledist}. We show that if the error is a small constant $\approx \frac{1}{|C|^4}$, then the only strategy Alice can employ is to send the register $C$ to Bob. For this, we view $\cF$ as a transition matrix for probability distributions and show that it must be close to the identity matrix.  We obtain the following. 
\begin{theorem}
\label{thm:infmain}
The following holds for the ensemble of two equiprobable states
$\{\rho^0_C,\rho^1_C\}$, where $|C| = d$, $\rho^0_C = I/d$, $\rho^1_C$ is
diagonal, with $(c,c)$-entry being $(d-c+1)/\eta$, and $\eta = d(d+1)/2$.  
The achievable rate for entanglement-assisted blind  
compression is at least $(\log d) - 7$ bits, while the Holevo information
$\mutinf{X}{C}_{\rho}$ is at most $1$.  
The lower bound holds for both global and local errors
of $\eps\approx \frac{1}{d^4}$, which is independent of
the number of instances $n$.
Thus the information defect at non-vanishing local error can be 
arbitrarily large, and near maximal for the dimension.
\end{theorem}

Our proof highlights a `strong no-cloning principle' in the classical
setting. To clarify, observe that Alice and Bob cannot 
transform (or
clone) $\rho^x_C \rightarrow \rho^x_C\otimes \rho^x_{C'}$ without the
knowledge of $x$. This translates to the statement that
$\min_{\cF:C\rightarrow CC'} \condmutinf{C}{C'}{X}_{\cF(\rho)}$ is
bounded away from $0$ in (\ref{eq:pinskerlb})(c). Theorem \ref{thm:infmain}
goes further to show that the only way to produce the register $C'$ 
creates a lot of correlation between $C$ and $C'$. This is akin to
the situation in quantum no-cloning; the operation
$\ket{i}_C\rightarrow \ket{i}_C\ket{i}_{C'}$ leads to a large
correlation when applied to a state $\sum_i \alpha_i\ket{i}_C$ in
superposition.

\subsection{Conclusion}

In this work, we study the problem of blind compression of quantum
data, in the regime of finite error. Our inspiration comes from two
sources. First is the visible scenario, where the trade-off between
global error and communication rate is very well understood (providing
a strong converse rate) and the trade-off between local error and
communication rate is relatively well understood. Second is the
Koashi-Imoto characterization, which gives the optimal rate of
communication as the error vanishes in the asymptotic limit and hence
shows a near maximal separation between the communication rate and the
Holevo information in the vanishing error regime.  
\change{In their converse bound for non-vanishing error, the error
dependence of the rate is not explicitly given, and we have observed
sensitivity to error in the appendix.
Thus the Koashi-Imoto characterization does not immediately 
apply to the current regime of non-vanishing (global or local) error. }
Our main result resolves this
problem, showing a near maximal separation of the rate  
from the Holevo information
in the non-vanishing local error regime. For this, we prove a new
lower bound that is based on a variant of the no-cloning theorem for
classical distributions.  Our technical proof builds on an approximate
version of the Birkoff-von Neumann theorem.

An immediate question raised by our work is to understand the error
vs communication rate trade-off for the blind compression scenario, for
the cases of global and local errors. Furthermore, we ask if a strong
converse rate exists for the blind compression scenario when the
global error is finite, which is known to hold for the visible
case. Finally, we highlight that our lower bound does not entirely
rely on the spatial separation between the sender and the receiver,
which leads to the question of further applicability of our techniques
to other problems.

\section{Notations and information theoretic quantities used}
\label{sec:prelims}

\subsection{Basic notions in quantum information theory}
Throughout the paper, $\log$ is taken base 2.  
For a finite set $\cC$, a probability distribution is a function $p:\cC \rightarrow [0,1]$ satisfying $\sum_{c\in \cC}p(c)=1$.
In this paper, we only consider finite dimensional Hilbert spaces.  
Consider such a Hilbert space $\cH$ endowed with an inner product $\langle \cdot, \cdot \rangle$.  
For an operator $X$ acting on $\cH$, the Schatten-$1$
norm of $X$ is defined as $\| X\|_1:=\Tr\sqrt{X^{\dagger}X}$ and the
Schatten-$2$ norm is defined as $\| X\|_2:=\sqrt{\Tr XX^{\dagger}}$. A quantum state is represented by a density matrix $\rho$, which is a positive semi-definite operator on $\cH$ with trace equal to $1$. The quantum state $\rho$ is {\em pure} if and only if its density matrix is rank $1$, in which case $\rho = \ketbra{\psi}$ for some unit vector $\ket{\psi} \in \cH$.
Throughout the paper, we may use $\psi$ to represent the quantum state and also the density matrix $\ketbra{\psi}$. Given a quantum state $\rho$ on $\cH$, {\em the support of $\rho$}, denoted $\text{supp}(\rho)$, is the subspace of $\cH$ spanned by all eigenvectors of $\rho$ with \emph{positive} eigenvalues.

A {\em quantum register} $A$ is associated with some Hilbert space $\cH_A$. Define $|A| := \dim(\cH_A)$. Let $\mathcal{L}(A)$ represent the set of all linear operators on $\cH_A$.
We denote the set of quantum states on the Hilbert space $\cH_A$ by $\mathcal{D}(A)$. The quantum state $\rho$ with subscript $A$ indicates $\rho_A \in \mathcal{D}(A)$. If two registers $A,B$ are associated with isomorphic Hilbert spaces (that is, of the same dimension), we write $A\equiv B$. Two disjoint registers $A$ and $B$ combined, denoted as $AB$, is associated with the tensor product Hilbert space $\cH_A \otimes \cH_B$.
For two operators $M_1 \in \mathcal{L}(A)$ and $M_2 \in \mathcal{L}(B)$, $M_1 \otimes M_2 \in \mathcal{L}(AB)$ represents the tensor product (Kronecker product) of $M_1$ and $M_2$. 
The identity operator on $\cH_A$ (and its associated register $A$) is denoted as $\id_A$.

For any operator $M_{AB} \in \mathcal{L}(AB)$, the partial trace on $A$ is defined as:
\[ \Tr_{A} \, (M_{AB})
:= \sum_i (\bra{i} \otimes \id_{B})
\, M_{AB} \, (\ket{i} \otimes \id_{B}) , \]
where $\{\ket{i}\}_i$ is an orthonormal basis for the Hilbert space $\cH_A$.
For a quantum state $\rho_{AB} \in \mathcal{D}(AB)$,
we use the shorthand
$$\rho_{B} := \Tr_{A} \, (\rho_{AB})$$  
and the quantum state $\rho_B \in \mathcal{D}(B)$ is referred to as the marginal quantum state of $\rho_{AB}$. Unless otherwise stated, a missing register from the subscript of a quantum state represents a partial trace over that register.

A  quantum state  $\rho_{AB}$ is  \textit{classical-quantum} with  $A$
being the classical register, if it  is of the form $\rho_{AB}= \sum_a
p_A(a)\ketbra{a}\otimes  \rho^a_B$,  where  $\{\ket{a}\}_a$  forms  an
orthonormal basis,  $\{p_A(a)\}_a$ is  a probability  distribution and
$\rho^a_B\in  \mathcal{D}(B)$.  The value $a$ stored in register $A$ 
identifies a corresponding quantum state $\rho^a_B$ on register $B$.
This convention allows a clear distinction between having a specimen
of the state $\rho^a_B$ (having system $B$) and knowing what it is
(having system $A$).  If all $\rho^a_B$ are diagonal in the same basis, 
$\rho_{AB}$ is called \textit{classical-classical} or simply \textit{classical}.

A quantum {channel} $\cE: \mathcal{L}(A)\rightarrow \mathcal{L}(B)$ is a completely positive and trace preserving (CPTP) linear map.  (We sometimes just call it a ``map'' in this paper.)  
In particular, it takes quantum states in $\mathcal{D}(A)$ to the quantum states in $\mathcal{D}(B)$.
A quantum measurement (or instrument) $\cN: \cL(A) \rightarrow \cL(A'C)$ is characterized by a collection of operators $\{N_c : \cH_A \rightarrow \cH_{A'}\}$ that satisfy $\sum_c N_c^{\dagger}N_c = \id_A$ and is given by
$$\cN(\rho_A) = \sum_c \ketbra{c}_C\otimes N_c\rho_AN^{\dagger}_c.$$
 A {\em unitary} operator $U_A:\cH_A \rightarrow \cH_A$ is such that $U_A^{\dagger}U_A = U_A U_A^{\dagger} = \id_A$. 

\subsection{List of quantum information theoretic quantities}

We consider the following information theoretic quantities. All logarithms are base $2$ and only normalized quantum states are considered in the definitions below. Let $\varepsilon \in (0,1)$. 
\begin{enumerate}
\item {\bf Trace distance:}. For $\rho_A,\sigma_A \in \mathcal{D}(A)$, $$\Delta(\rho_A,\sigma_A) := \frac{1}{2}\|\rho_A - \sigma_A\|_1.$$
\item {\bf Fidelity:} For $\rho_A,\sigma_A \in \mathcal{D}(A)$, $$\F(\rho_A,\sigma_A) := \|\sqrt{\rho_A}\sqrt{\sigma_A}\|_1.$$
\item {\bf $\varepsilon$-ball:} For $\rho_A\in \mathcal{D}(A)$, $$\ball{\eps}{\rho_A} := \{\rho'_A\in \mathcal{D}(A)|~\Delta(\rho_A,\rho'_A) \leq \varepsilon\}. $$ 

\item {\bf Von Neumann entropy:} (\cite{Neumann32}) For $\rho_A\in\mathcal{D}(A)$, $$\ent{A}_{\rho} := - \Tr(\rho_A\log\rho_A) .$$ 
\item {\bf Conditional entropy:}  For $\rho_{AB}\in\mathcal{D}(AB)$, $$\cent{A}{B}_{\rho} := \ent{AB}_{\rho} - \ent{B}_{\rho} .$$ 
\item {\bf Relative entropy:} (\cite{umegaki1954}) For $\rho_A,\sigma_A\in \mathcal{D}(A)$ such that $\text{supp}(\rho_A) \subset \text{supp}(\sigma_A)$, $$\relent{\rho_A}{\sigma_A} := \Tr(\rho_A\log\rho_A) - \Tr(\rho_A\log\sigma_A) .$$ 
\item {\bf Mutual information:} For $\rho_{AB}\in \mathcal{D}(AB)$, $$\mutinf{A}{B}_{\rho}:= \ent{A}_{\rho} + \ent{B}_{\rho}-\ent{AB}_{\rho}= \relent{\rho_{AB}}{\rho_A\otimes\rho_B}.$$
\item {\bf Conditional mutual information:} For $\rho_{ABC}\in\mathcal{D}(ABC)$, $$\condmutinf{A}{B}{C}_{\rho} := \mutinf{A}{BC}_{\rho}-\mutinf{A}{C}_{\rho}.$$
\end{enumerate}

\subsection{Basic facts used in our proofs}

\begin{fact}[Triangle inequality for trace distance,~\cite{NielsenC00}, Chapter 9]
\label{fact:trianglepurified}
For quantum states $\rho, \sigma, \tau\in \mathcal{D}(A)$,
$$\Delta(\rho,\sigma) \leq \Delta(\rho,\tau)  + \Delta(\tau,\sigma) . $$ 
\end{fact}

\begin{fact}[Data-processing inequality, \cite{barnum96,lindblad75}]
	\label{fact:dataproc}
For the quantum states $\rho$, $\sigma \in \mathcal{D}(A)$, $\theta \in \mathcal{D}(AC)$ and the quantum channel $\cE:\mathcal{L}(A)\rightarrow \mathcal{L}(B)$, it holds that
\begin{align*}
\Delta(\rho,\sigma) & \; \change{\geq} \; \Delta(\cE(\rho),\cE(\sigma)),\\  \relent{\rho}{\sigma}&\geq \relent{\cE(\rho)}{\cE(\sigma)}, \\
\mutinf{A}{C}_{\theta}  & \geq \mutinf{B}{C}_{\cE (\theta)}.
\end{align*}
\end{fact}

\begin{fact}[Pinsker's inequality, \cite{Pinsker64}]
\label{pinsker}
For the quantum states $\rho,\sigma\in\mathcal{D}(A)$, 
$$\Delta(\rho,\sigma)^2 \leq \frac{1}{2}\relent{\rho}{\sigma}.$$
\end{fact}

\begin{fact}[Dimension bound]
	\label{fact:dimbound}
For the quantum state $\rho_{XAB}$, with classical register $X$, it holds that
\begin{align*}
\condmutinf{A}{X}{B}_{\rho}  & \leq \log|X|.
\end{align*}
\end{fact}

\begin{fact}[Alicki-Fannes-Winter inequality, \cite{alicki2004continuity, winter2016tight}]
\label{alickifannes}
For quantum-classical states $\rho_{AB}$ and $\sigma_{AB}$ satisfying~$\rho_{B} = \sigma_{B}$,
$$|\cent{A}{B}_{\rho}-\cent{A}{B}_{\sigma}| \leq \Delta\br{\rho_{AB},\sigma_{AB}}\cdot\log|A|+ 1,$$
$$|\mutinf{A}{B}_{\rho}-\mutinf{A}{B}_{\sigma}| \leq \Delta\br{\rho_{AB},\sigma_{AB}}\cdot \log|B| + 1.$$
\end{fact}


\begin{fact}[Fano's inequality, \cite{Fano61}]
\label{fano}
For any classical state $\rho_{AA'} = \sum_{a,a'}p_{AA'}(a,a')\ketbra{a,a'}$, with $p_{AA'}$ a probability distribution, it holds that
$$\cent{A}{A'}_{\rho}\leq 1+ \Pr[A\neq A']\log|A|.$$
\end{fact}

\noindent Note that we have stated weaker 
versions of Alicki-Fannes-Winter inequality and Fano's inequality that
simplify the expressions in our results.

\section{Lower bound on entanglement-assisted blind distribution compression}
\label{sec:singlelettbdd}

\noindent For our lower bound on the compression rate, we focus on ensembles of \emph{classical} states (these can be simultaneously diagonalized).  We will henceforth refer to them as distributions.  We begin with a formal definition of our task.  
\begin{definition}[Entanglement-assisted blind distribution compression]
\label{def:mixedcomp}
Consider an ensemble $\{p(x), \rho^x_C\}$ where all $\rho^x_C$'s are diagonal.   
Let $\rho_{XC}$ be as \change{obtained through} (\ref{rhoxc}).  Let $\eps\in (0,1)$ be an error parameter and $n \in \bbN$. Let the initial joint state between the Referee and Alice be $\rho_{X_1C_1}\otimes \rho_{X_2C_2}\otimes \ldots \otimes \rho_{X_nC_n}$, with the Referee holding registers $X_1, X_2, \ldots, X_n$ (each $X_i\equiv X$) and Alice holding registers $C_1, C_2, \ldots, C_n$ (each $C_i\equiv C$).  Alice and Bob share entanglement $\ket{\theta}_{E_AE_B}$, where $E_A$ is with Alice and $E_B$ is with Bob. An $(n, R, \eps)$- entanglement-assisted blind distribution compression protocol is as follows. Alice applies an encoding map $\cE: C_1C_2\ldots C_nE_A\rightarrow C_1C_2\ldots C_nT$, where $T$ is a classical register of size $2^{nR}$. She communicates $T$ to Bob (so the number of bits communicated in the protocol is $nR = \log|T|$).  After receiving $T$, Bob applies a decoding map $\cD: TE_B\rightarrow C'_1C'_2\ldots C'_n$. Here, each $C'_i \equiv C$. It is required that
\beq
\label{eq:errorcond}
\Delta\br{\Tr_{C_1 \ldots C_n} \circ \cD\circ\cE\br{\rho_{X_1C_1}
\otimes \ldots \otimes  \rho_{X_nC_n}\otimes \theta_{E_AE_B}}, \rho_{X_1C'_1}
\otimes \ldots \otimes \rho_{X_nC'_n}}\leq \eps.
\enq
The above definition involves a global error for the compression.  Our lower bounds apply also to the more relaxed setting of the local error model:
\beq
\label{eq:locerror}
\forall_i \; \Delta\br{\Tr_{C'_1 \ldots C'_{i-1} C'_{i+1} \ldots C'_n C_1 \ldots C_n} \circ \cD\circ\cE
\left( \rho_{X_1C_1} \otimes 
\ldots \otimes  \rho_{X_nC_n}\otimes \theta_{E_AE_B} \right), \rho_{X_iC'_i}} \leq \eps.
\enq
\end{definition}

\noindent Note that the definition uses classical communication, which is equivalent to quantum communication up to a factor of $2$ when entanglement is free.

Since the ensemble is classical, Alice can always retain the information in the registers $C_1, C_2, \ldots, C_n$, so, without loss of generality, we assume the following equality throughout the discussion.
\beq
\label{eq:classicalcopy}
\Tr_T\br{\cE\br{\rho_{X_1C_1}\otimes \rho_{X_2C_2}\otimes \ldots \otimes \rho_{X_nC_n}\otimes \theta_{E_A}}}= \rho_{X_1C_1}\otimes \rho_{X_2C_2}\otimes \ldots  \otimes \rho_{X_nC_n}. 
\enq
The following theorem shows a lower bound on the rate of communication $R$
required for the task.
\begin{theorem}
\label{thm:mixedlb}
Let $\rho_{XC}$ be as \change{given} in Definition \ref{def:mixedcomp}, $n$ a
natural number, and $\eps\in (0,1)$. For any $(n, R, \eps)$-
entanglement-assisted blind distribution compression, it holds that
$$R \; \geq \min_{\cF:C\rightarrow CC'}\br{\condmutinf{C}{C'}{X}_{\tau}}+\mutinf{X}{C}_{\rho} - \eps\log|X|-1,$$
where \change{$\tau_{XCC'}:=\cF(\rho_{XC})$ and} the map $\cF$ must satisfy \change{$\Delta\br{\tau_{XC'}, \rho_{XC'}}\leq \eps$ and $\tau_{XC}=\rho_{XC}$}.
\end{theorem}
\begin{proof}
For brevity, set $\bX^n = X_1X_2\ldots X_n$, $\bC^n = C_1C_2\ldots C_n$ and $\bC'^n = C'_1C'_2\ldots C'_n$.
Let
$\sigma_{\bX^n\bC^nTE_B}$ be the state after Alice's encoding, 
and $\tau_{\bX^n\bC^n\bC'^n}$ be the final quantum state at the end of the protocol.
Observe that
\beq
\label{eq:2nRlb}
nR \; = \; \log|T| 
\; \geq \; \condmutinf{\bX^n\bC^n}{T}{E_B}_\sigma \; \overset{(a)}= \; \mutinf{\bX^n\bC^n}{TE_B}_\sigma \; \overset{(b)}\geq \; \mutinf{\bX^n\bC^n}{\bC'^n}_\tau. 
\enq
The equality $(a)$ in (\ref{eq:2nRlb}) follows from the fact that
$\mutinf{\bX^n\bC^n}{E_B}_{\sigma}=0$.  We apply the data processing
inequality to obtain $(b)$.  Note also from this step onwards, 
entanglement no longer plays a role in the proof.  Now, consider
\beqar
\label{eq:XCClb}
\mutinf{\bX^n\bC^n}{\bC'^n}_\tau &=& \sum_{i=1}^n \condmutinf{X_iC_i}{\bC'^n}{X_1\dots X_{i-1}C_1\ldots C_{i-1}}_{\tau}\nonumber\\
&=& \sum_{i=1}^n \br{\mutinf{X_iC_i}{\bC'^nX_1\dots X_{i-1}C_1\ldots C_{i-1}}_{\tau}- \mutinf{X_iC_i}{X_1\dots X_{i-1}C_1\ldots C_{i-1}}_{\tau}}\nonumber\\
&\overset{(c)}=& \sum_{i=1}^n \mutinf{X_iC_i}{\bC'^nX_1\dots X_{i-1}C_1\ldots C_{i-1}}_{\tau}\nonumber\\
&\geq& \sum_{i=1}^n \mutinf{X_iC_i}{C'_i}_{\tau}.
\enqar
In (\ref{eq:XCClb}), the equality $(c)$ holds since (\ref{eq:classicalcopy}) ensures that $$\tau_{X_iC_iX_1\dots X_{i-1}C_1\ldots C_{i-1}}=\rho_{X_iC_iX_1\dots X_{i-1}C_1\ldots C_{i-1}}=\rho_{X_iC_i}\otimes\rho_{X_1\dots X_{i-1}C_1\ldots C_{i-1}},$$ 
and the last step follows from the data processing inequality.  
From (\ref{eq:errorcond}), we have $\Delta\br{\tau_{X_iC'_i}, \rho_{X_iC'_i}} \leq \eps$. Thus, using \change{the Alicki-Fannes-Winter inequality (Fact \ref{alickifannes}),} we obtain
\beq
\label{eq:XiCilb}
\mutinf{X_iC_i}{C'_i}_{\tau} = \condmutinf{C_i}{C'_i}{X_i}_{\tau}+\mutinf{X_i}{C'_i}_{\tau}\geq \condmutinf{C_i}{C'_i}{X_i}_{\tau}+\mutinf{X_i}{C_i}_{\rho} -\eps\log|X|-1.
\enq
Combining (\ref{eq:2nRlb})-(\ref{eq:XiCilb}),
we obtain
$$nR\geq n\min_i\br{\condmutinf{C_i}{C'_i}{X_i}_{\tau}+\mutinf{X_i}{C_i}_{\rho} -\eps\log|X|-1}.$$
We can now convert the above asymptotic inequality to a single-letter bound.  
For an $i$ that achieves the minimum, define $\cF_i$ to be the map that acts on register $C_i$ as follows. It first creates the state $\rho_{X_1 C_1}\otimes \ldots \otimes  \rho_{X_{i-1} C_{i-1}}\otimes \rho_{X_{i+1} C_{i+1}}\otimes\ldots \otimes \rho_{X_n C_n}\otimes \theta_{E_AE_B}$. Then it applies $\cD\circ\cE$ and traces out registers $X_1, \ldots , X_{i-1}, X_{i+1},\ldots , X_n$,  $C_1,\ldots, C_{i-1}, C_{i+1},\ldots, C_n$, $C'_1, \ldots,  C'_{i-1}, C'_{i+1}, \ldots, C'_n$. From (\ref{eq:errorcond}), we conclude that
$$\Delta\br{\Tr_{C_i}\cF_i(\rho_{XC_i}), \rho_{XC'_i}} \leq \eps.$$
Moreover, $\Tr_{C'_i}\cF(\rho_{XC_i})=\rho_{XC_i}$, as the maps $\cE$ and $\cD$ do not change the state in registers $XC_i$. Since $C_i\equiv C$, $C'_i\equiv C'$ and $\rho_{XC_i}=\rho_{XC}$, the proof concludes.   
\end{proof}

Theorem \ref{thm:mixedlb} shows that the communication cost for entanglement-assisted blind mixed distribution compression can exceed the Holevo information of the distribution $\mutinf{X}{C}_{\rho}$.
We now show that this additional cost \change{of $R-\mutinf{X}{C}_{\rho}$} can be quantitatively bounded by some measure of indistinguishability of the states in the ensemble, and also by some measure of the inability to clone the states.  
To proceed with this, consider a simple example of compressing two equiprobable distributions, with
$\rho_{XC} = \frac{1}{2} (\ketbra{0}_X \otimes \rho_C^0 + \ketbra{1}_X \otimes \rho_C^1)$.  For a map $\cF$ satisfying $\Delta(\Tr_{C}\br{\cF(\rho_{XC})}, \rho_{XC'})\leq \eps$ and $\Tr_{C'}\cF(\rho_{XC})=\rho_{XC}$, let $\tau_{XCC'} = \cF(\rho_{XC})$.  We will prove the following: 
\beqar
\label{eq:pinskerlb}
\sqrt{\condmutinf{C}{C'}{X}_{\cF(\rho_{XC})}}&\overset{(a)} = & \sqrt{\frac{1}{2} \left( D(\tau^0_{CC'} || \tau^0_C\otimes \tau^0_{C'}) + D(\tau^1_{CC'} || \tau^1_C\otimes \tau^1_{C'}) \right) }\nonumber\\
&\overset{(b)}\geq & \sqrt{\Delta(\tau^0_{CC'},\tau^0_C\otimes \tau^0_{C'})^2 + \Delta(\tau^1_{CC'},\tau^1_C\otimes \tau^1_{C'})^2}\nonumber\\
&\overset{(c)}\geq& \frac{1}{\sqrt{2}}\left( \Delta(\tau^0_{CC'},\tau^0_C\otimes \tau^0_{C'}) + \Delta(\tau^1_{CC'},\tau^1_C\otimes \tau^1_{C'}) \right) \nonumber\\
&\overset{(d)}\geq& \frac{1}{\sqrt{2}}\br{\Delta(\tau^0_C\otimes \tau^0_{C'},\tau^1_C\otimes \tau^1_{C'})-\Delta(\tau^0_{CC'},\tau^1_{CC'})}\nonumber\\
&\overset{(e)}\geq& \frac{1}{\sqrt{2}}\br{\Delta(\rho^0_C\otimes \rho^0_{C'},\rho^1_C\otimes \rho^1_{C'})-\Delta(\tau^0_{CC'},\tau^1_{CC'})-2\eps}\nonumber\\
&\overset{(f)}\geq& \frac{1}{\sqrt{2}}\br{\Delta(\rho^0_C\otimes \rho^0_{C'},\rho^1_C\otimes \rho^1_{C'})-\Delta(\rho^0_{C},\rho^1_{C})-2\eps}.
\enqar
Here, $(a)$ uses the expansion 
$$\condmutinf{C}{C'}{X}_{\cF(\rho)}= \condmutinf{C}{C'}{X}_{\tau}= \sum_x \, p(x) \, \relent{\tau^x_{CC'}}{\tau^x_C\otimes \tau^x_{C'}},$$
and $(b)$ uses Pinsker's inequality
(Fact \ref{pinsker}), 
$(c)$ follows from the inequality $a^2+b^2 \geq \frac{(a+b)^2}{2}$, $(d)$ uses the triangle inequality for trace distance, $(e)$ uses the identity $\tau^x_C=\rho^x_C$ and the inequality $\Delta(\tau_{XC'}, \rho_{XC'}) =  \frac{1}{2} \Delta(\tau^0_{C'}, \rho^0_{C'}) + \frac{1}{2} \Delta(\tau^1_{C'}, \rho^1_{C'}) \leq \eps$ and $(f)$ uses the data-processing inequality (Fact \ref{fact:dataproc}) to conclude that $\Delta(\tau^0_{CC'},\tau^1_{CC'})=\Delta \left( \cF\br{\rho^0_{C}},\cF\br{\rho^1_{C}}) \leq \Delta(\rho^0_{C},\rho^1_{C} \right)$. 

Furthermore, the above chain of inequalities quantitatively relate the gap between the communication cost and the Holevo information to other quantitative properties of the ensemble.  Recall that $\tau^x_C=\rho^x_C$, so the RHS of the inequality (c) lower-bounds the gap by a ``classical no-cloning bound'', which is the average distance between two copies of $\rho^x_C$ and the actual Alice-Bob joint-output.  Furthermore, the RHS of the inequality (f) says that the gap is lower-bounded by the \emph{increase} in \change{distinguishability} of $\rho_C^0$ and $\rho_C^1$ if a second copy is available, which is a measure of the indistinguishability between $\rho_C^{0,1}$.  

This gap can be strictly positive for some ensemble.  For example, consider:
$$\rho^0_C=\begin{pmatrix} \frac{1}{2} & 0 \\ 0 & \frac{1}{2} \end{pmatrix}, \quad \rho^1_C=\begin{pmatrix} \frac{1}{3} & 0 \\ 0 & \frac{2}{3} \end{pmatrix}.$$
We evaluate 
$$\Delta(\rho^0_{C},\rho^1_{C}) =\frac{1}{2}\br{\bigg|\frac{1}{2}-\frac{1}{3}\bigg| + \bigg|\frac{1}{2}-\frac{2}{3}\bigg|} = \frac{1}{6}$$
and
$$\Delta(\rho^0_C\otimes \rho^0_{C},\rho^1_C\otimes \rho^1_{C})= \frac{1}{2}\br{\bigg|\frac{1}{4}-\frac{1}{9}\bigg| + 2\bigg|\frac{1}{4}-\frac{2}{9}\bigg|+\bigg|\frac{1}{4}-\frac{4}{9}\bigg|} = \frac{7}{36}.$$
We conclude that $\Delta(\rho^0_C\otimes \rho^0_{C},\rho^1_C\otimes \rho^1_{C})-\Delta(\rho^0_{C},\rho^1_{C}) = \frac{1}{36}$. Thus, $$\condmutinf{C}{C'}{X}_{\cF(\rho)} \geq \frac{(1-72\eps)^2}{2\cdot 6^4}.$$

\noindent 
This example demonstrates a constant lower bound on the information defect.  
In the next section, we develop a large lower bound that 
is nearly maximal given the dimension of $C$.

\section{Near maximal separation between information cost and the communication cost}

\change{In this section, we prove Theorem \ref{thm:infmain}. The proof will proceed in the following steps.}

\begin{itemize}
\item \change{We will simplify the form of the map $\cF$ by observing that the input and the output are classical. This will also lead to a simpler lower bound than given in Theorem \ref{thm:mixedlb}, in terms of stochastic maps. }
\item \change{We will show the utility of the simpler lower bound first in the case of zero error. We will show that the stochastic map under consideration is actually doubly stochastic, since one of the two distributions in the ensemble is uniform. Then we apply the Birkoff-von Neumann theorem to show that the only relevant doubly stochastic map is the identity map, which implies our lower bound.}
\item \change{Finally, we will consider the case of non-zero error. We will show that the stochastic map under consideration is close to a doubly stochastic map. Then we will analyze the action of this doubly stochastic map on the distributions in the ensemble. Since the doubly stochastic map keeps the distributions approximately unchanged, we bound its distance from the identity map. This will lead to the desired lower bound for the communication cost.}
\end{itemize}

\subsection{Stochastic maps and the distribution}

\change{Recall the state $\rho_{XC}$, the map $\cF$ and the state $\tau_{XCC'}$ in Theorem \ref{thm:mixedlb}}. We start with the observation that, despite the fact that the map $\cF$ involves entanglement, the state $\tau_{XCC'}=\cF\br{\rho_{XC}}$ can be considered to be completely classical, without loss of generality. To see this, consider the constraints $$\Delta(\Tr_{C}\br{\cF(\rho_{XC})}, \rho_{XC'})\leq\eps$$ and $\Tr_{C'}\cF(\rho_{XC})=\rho_{XC}$ \change{as in Theorem \ref{thm:mixedlb}}. Let $\cF'$ be the map resulting from measuring the register $C'$ after applying $\cF$ to $\rho_{XC}$. Then it still holds that  $$\Delta(\Tr_{C}\br{\cF'(\rho_{XC})}, \rho_{XC'})\leq\Delta(\Tr_{C}\br{\cF(\rho_{XC})}, \rho_{XC'})\leq \eps$$ and $\Tr_{C'}\cF'(\rho_{XC})=\rho_{XC}$, so the constraints are still satisfied, and moreover
$$\condmutinf{C}{C'}{X}_{\cF'(\rho)}\leq\condmutinf{C}{C'}{X}_{\cF(\rho)}$$
by the data processing inequality.
Therefore \change{in Theorem \ref{thm:mixedlb}}, it suffices to \change{restrict to} maps $\cF$ with completely classical $\tau_{XCC'}$, which has the form 
\begin{equation}
\label{eq:deftau}
\tau_{XCC'}=\sum_{x,c,c'}p_{XCC'}(x,c,c')\ketbra{x,c,c'}.
\end{equation}
Note that $p_{XCC'}(x,c,c')$ satisfies the Markov chain condition $p_{XC'|C=c}(x,c')=p_{X|C=c}(x)\,p_{C'|C=c}(c')$, as the map $\cF$ produces $C'$ from $C$.  Define a matrix $M$ with 
\begin{equation}
\label{eq:defM}
M_{c,c'} = p_{C'|C=c}(c').
\end{equation}
Clearly, $M$ is a stochastic matrix: $\forall_c \, \sum_{c'} M_{c,c'} = 1$.
\change{Therefore each distribution on $C$ is mapped to a distribution on $C'$.} 
\change{The rest of the paper revolves about understanding this stochastic matrix $M$.  }
First, we view $M$ as a `channel' between Alice and Bob that inputs $C$ and outputs $C'$, and appeal to Fano's inequality (Fact \ref{fano}) to bound the information between $C$ and $C'$ in terms of the probability of $C$ being equal to $C'$. This is shown in the following claim.
\begin{claim}
\label{clm:genlowb}
\change{Given the states $\rho$, $\tau$ and the associated stochastic matrix $M$ defined above,} it holds that
$$\condmutinf{C}{C'}{X}_\tau \geq \cent{C}{X}_{\rho} -2 -\br{1-\sum_cp_C(c)M_{c,c} + \eps}\log\br{|C|}.$$
\end{claim}

\begin{proof}
Using the definition for conditional entropy, the Markov chain property 
$\cent{C'}{CX}_{\tau} = \cent{C'}{C}_{\tau}$ on $\tau$, 
and applying the Alicki-Fannes-Winter inequality along with the condition 
$\Delta(\Tr_{C}\br{\cF(\rho_{XC})}, \rho_{XC'})\leq\eps$ to 
$\cent{C'}{X}_\tau$, we obtain
\beqar
\condmutinf{C}{C'}{X}_\tau &=& \cent{C'}{X}_{\tau} - \cent{C'}{CX}_{\tau} \nonumber\\
&=& \cent{C'}{X}_{\tau} - \cent{C'}{C}_{\tau} \nonumber 
\\
&\geq& \cent{C'}{X}_{\rho} - \cent{C'}{C}_{\tau} - \Delta(\rho_{XC'}, \tau_{XC'}) \log(|C|) - 1 \nonumber
\\
&\geq& \cent{C}{X}_{\rho} - \cent{C'}{C}_{\tau} - \eps \log(|C|) - 1.
\enqar
Using Fano's inequality (Fact \ref{fano})
\beqarst
\cent{C'}{C}_{\tau}&\leq& 1+ \Pr[C\neq C']\log\br{|C|} = 1+ \br{1- \sum_cp_C(c)M_{c,c}}\log\br{|C|}.
\enqarst
This concludes the proof.
\end{proof}

\noindent {\bf The ensemble.} Using Claim \ref{clm:genlowb}, our strategy to obtain a large gap between the communication cost for compression and the Holevo information is to construct ensembles for which $\sum_cp_C(c)M_{c,c} \approx 1$, which implies $\condmutinf{C}{C'}{X}_\tau \approx \cent{C}{X}_{\rho}$.  
We now define our ensemble using the following notations.  
Let $d = |C|$, $c\in \{1,2,\ldots d\}$, $x\in\{0,1\}$, and $p_X(0)=p_X(1)=\frac{1}{2}$.  We choose the states $\rho^0_C$ and $\rho^1_C$ to correspond to the uniform and `staircase' distributions (see Figure \ref{fig:exampledist} \change{and also Theorem \ref{thm:infmain}}), defined as follows.
\beq
\label{eq:ourdist}
p_{C|X=0}(c) = \frac{1}{d} \; {=}{:} \; u_c \,, \quad p_{C|X=1}(c) = \frac{d-c+1}{\eta} \; {=}{:} \; v_c \,,
\enq
where $\eta=\frac{d(d+1)}{2}$.  We use $u$ to denote the row vector with the $c$-th entry being $u_c$, and similarly for $v$.

\subsection{Lower bound in the zero error case}

To illustrate the principles in the general case, we 
first consider the zero-error case where $\eps=0$. \change{In this case, we have $\tau_{XC'}=\rho_{XC'}$. From the definition of the stochastic matrix $M$ in Equation \ref{eq:defM} and the choice of distributions $p_{C|X=0}, p_{C|X=1}$ given above, we arrive at the following set of equations.}
\beq
\label{fixedp}
u M = u, vM = v \,, 
\enq
so both $u, v$ are fixed points of the transition matrix $M$. We have the following theorem.
\begin{theorem}
\label{thm:zeroerror}
Using aforementioned definitions, (\ref{fixedp}) implies that 
$M_{c,c'} =1 \text{ iff } c'=c$ (i.e., $M=\id$).  
\end{theorem}
\begin{proof}
Using the conditions $u M = u$ and $u_c=\frac{1}{|C|}$ for all $c$, (\ref{fixedp}) implies that  
$$\forall_{c'} \; \sum_c M_{c,c'}=1.$$
Thus the matrix $M$ is doubly stochastic. From the Birkoff-von Neumann theorem \cite{Konig36, Birkoff46, Neumann53},   
there exist permutation matrices 
$\Pi_1,\Pi_2,\ldots \Pi_k$ and a probability distribution $(q_1, \ldots q_k)$ such that
\beq
\label{eq:birkoffvN}
M=\sum_i q_i\Pi_i.
\enq
Next, we show that $M=\id$.
Without loss of generality, $\Pi_1=\id$, $0 \leq q_1 \leq 1$.  
Suppose, by contradiction, $q_1 < 1$, so there exists $i \geq 2$ 
with $q_i > 0$ and $\Pi_i \neq \id$.  Using $vM=v$ and 
applying (\ref{eq:birkoffvN}), 
\beqst
\sum_i q_i \br{v\Pi_i} = v.
\enqst 
Since $v\Pi_1=v$, we obtain 
\beq
\label{convcomb}
\frac{1}{1-q_1}\sum_{i=2}^kq_i\br{v\Pi_i} = v.
\enq
We will argue that this is a contradiction for the vector $v$. A
permutation $\Pi_i$ is said to act ``non-trivially'' on an index 
$j \in [d]$ if $j$ is not invariant under $\Pi_i$.  
Let $j_0\in [d]$ be the smallest index such that at least one of the permutations in
the set $\{\Pi_i\}_{i=2}^k$ act non-trivially on $j_0$. Since $\Pi_i \neq \id$ for 
$i=2,3,\cdots,k$, such an index $j_0$ must
exist. This implies the following:
\begin{enumerate}
 \item $\br{v\Pi_i}_{j}=v_j$ for all $j<j_0$, that is, all the permutations act trivially on indices smaller than $j_0$.
\item $\br{v\Pi_i}_{j_0}\leq v_{j_0}$, as any permutation $\Pi_i$ swaps the element $v_{j_0}$ with a potentially smaller element $v_j$. 
\item There is a permutation $\Pi_{i_0}$ such that $\br{v\Pi_{i_0}}_{j_0}< v_{j_0}$, by the definition of $j_0$.
\end{enumerate}
But items $2,3$ jointly contradict what is implied by the $j_0$-th entry of the vector equality in  (\ref{convcomb}):
$$\frac{1}{1-q_1}\sum_{i=2}^kq_i\br{v\Pi_{i}}_{j_0} = v_{j_0}.$$
Hence we must have $q_1=1$ and $M=\id$. This completes the proof.
\end{proof}

We conclude this subsection with a lower bound on the rate in the zero
error case.  When $\eps=0$, Theorem \ref{thm:zeroerror} implies that
$\sum_cp_C(c)M_{c,c}=1$.  Substituting this into Claim \ref{clm:genlowb}, 
$\condmutinf{C}{C'}{X}_{\tau} \geq S(C|X)_\rho - \eps \log(|C|)$.
Substituting this into Theorem \ref{thm:mixedlb}, 
$$R \geq \mutinf{X}{C}_{\rho} + S(C|X)_\rho - \eps (\log(|C|)+1) -1
\approx \mutinf{X}{C}_{\rho} + \alpha \log(|C|)$$
for some constant $\alpha$.  
In the next subsection, we proceed to the case of small error $\eps$.
We will generalize the intuition from the zero-error case in which $v$
is a vertex of the polytope formed by the convex hull of
$\{v\Pi_i\}_{i=1}^k$.


\subsection{Lower bound for the case of non-zero error}
In this case, \change{Theorem \ref{thm:mixedlb} ensures} that $\Delta\br{\tau_{XC'}, \rho_{XC'}}\leq \eps$. \change{Using the distribution $p_{XCC'}$ from Equation \ref{eq:deftau},} this translates to
$$\sum_xp_X(x)\sum_{c'} \, \Bigl| \, p_{C|X=x}(c')-p_{C'|X=x}(c') \Bigr| \leq 2\eps.$$
\change{For the rest of this paper, we continue to use the symbols $M, u, v$ defined earlier (see Equations \ref{eq:defM} and \ref{eq:ourdist}).}  
Translating the above \change{in terms} of $M, u, v$, 
and using the fact $p_X (0) = p_X (1) = \frac{1}{2}$,
$$\frac{1}{2} \bigl( \|u - uM\|_1 + \|v - vM \|_1 \bigr) \leq 2 \eps.$$
In particular, 
\beq
\label{eq:errorst}
\|u-uM\|_1 \leq 4\eps ~~{\rm and}~~
\|v-vM\|_1 \leq 4\eps.
\enq

We are ready to obtain our lower bound in spirits similar to the
zero-error case.  First, we will use the condition concerning $u$ in
(\ref{eq:errorst}) to approximate $M$ by a doubly
stochastic matrix $N$ (Lemma \ref{lem:appsto} part (a)).  Then we
approximate $vM$ by $vN$ (Lemma \ref{lem:appsto} part (b)).  With this
approximation, and with the condition on $v$ in (\ref{eq:errorst}),
we lower bound the identity component of $N$ which in turns lower
bound $N_{c,c}$ and then $M_{c,c}$ and the last bound gives us the
desired lower bound on the communication cost for the compression.  

\begin{lemma}
\label{lem:appsto}
Let $d \geq 2$ \change{and $M$ as defined before}. There exists a doubly stochastic matrix $N$ such that 
\begin{eqnarray}
{\bf {\rm (a)}} & \forall_{c,c'} \; |N_{c,c'}-M_{c,c'}| \leq 12d \eps\,,
\nonumber
\\[1ex] 
{\bf {\rm (b)}} & \|vN-vM\|_1\leq 12d \eps \,.
\end{eqnarray}
\end{lemma}

\begin{proof}
Rewriting the condition $\|u-uM\|_1 \leq 4\eps$ from (\ref{eq:errorst}) 
using $u_c = \frac{1}{d}$ for all $c$, 
\beq
\label{eq:errorst1}
\sum_{c'} \; \biggl| \sum_c M_{c,c'} -1 \biggr| \leq 4d\eps \,.  
\enq
Thus, the matrix $M$ is a stochastic matrix, but ``approximately'' a doubly 
stochastic matrix.  
In particular, define $\alpha_{c'} = \sum_c M_{c,c'} -1$ which measures 
how far $M$ deviates from being doubly stochastic.  
Using (\ref{eq:errorst1}), 
\beq
 \forall_{c'} ~~ | \alpha_{c'} | \leq 4d\eps \,.
\label{eq:errorst1-better}
\enq
We can now follow the idea from \cite{Khoury98} to find 
a doubly stochastic matrix $N$ that approximates $M$.  
Define 
$$N_{c,c'} = \frac{1}{1+4d\eps}\br{M_{c,c'} + \frac{4d\eps-\alpha_{c'}}{d}}.$$ 
First, note that the entries of $N_{c,c'}$ are non-negative, as 
$M_{c,c'}\geq 0$ and 
$4d\eps \geq \alpha_{c'}$ for all $c'$ (according to (\ref{eq:errorst1-better})). 
Second, $N$ is stochastic: 
\beqarst
\sum_{c'}N_{c,c'} &=& \frac{1}{1+4d\eps}\br{\sum_{c'}M_{c,c'} + \sum_{c'}\frac{4d\eps-\alpha_{c'}}{d}} \\ 
&=& \frac{1}{1+4d\eps}\br{1 + 4d\eps}=1,
\enqarst
where we have used the stochastic property of $M$ to substitute
$\sum_{c'} M_{c,c'}=1$, and also $\sum_{c'}\alpha_{c'}=\sum_{c,c'}
M_{c,c'} - d = 0$.
Next we show that $N$ is doubly stochastic.
\beqarst
\sum_{c}N_{c,c'} &=& \frac{1}{1+4d\eps}\br{\sum_{c}M_{c,c'} + \sum_{c}\frac{4d\eps-\alpha_{c'}}{d}} \\ 
&=& \frac{1}{1+4d\eps}\br{1+\alpha_{c'} + 4d\eps-\alpha_{c'}}=1.
\enqarst

Now, we bound the entry-wise difference between $M$ and $N$. Consider
\beqar
\label{eq:NminusM}
 N_{c,c'}- M_{c,c'} &=& \frac{1}{1+4d\eps}\br{M_{c,c'}+\frac{4d\eps - \alpha_{c'}}{d}}-M_{c,c'}\nonumber\\
&=& \frac{1}{1+4d\eps}\br{\frac{4d\eps- \alpha_{c'}}{d}-4d\eps M_{c,c'}}
\enqar
Thus,
\beqarst
| N_{c,c'}- M_{c,c'}| &\leq& \frac{1}{1+4d\eps}\br{\bigg|\frac{4d\eps- \alpha_{c'}}{d}\bigg|+4d\eps \bigg|M_{c,c'}\bigg|}\\
&\overset{(a)}\leq& 8\eps + 4d\eps \leq 12d\eps,
\enqarst
where $(a)$ uses $|M_{c,c'}|\leq 1$ and 
$|4d\eps - \alpha_{c'}| \leq 8d \eps$, in turns followed from 
(\ref{eq:errorst1-better}).  This proved claim (a).

For claim (b), define a matrix $W$ as
$W_{c,c'} = \frac{4d\eps- \alpha_{c'}}{d(1+4d\eps)}$. 
Equation (\ref{eq:NminusM}) states that 
$$N-M = W - \frac{4d\eps}{1+4d\eps} \,M \,.$$
Thus, 
\beqarst
\bigl\| vN-vM \bigr\|_1 &=& 
\left\| \; vW - \frac{4d\eps}{1+4d\eps} \; vM \; \right\|_1\\
&\leq& \left\| vW \right\|_1 + \frac{4d\eps}{1+4d\eps} \; \left\|vM \right\|_1\\
&\overset{(a)}=& \|vW\|_1 + \frac{4d\eps}{1+4d\eps}\\
&
=& \frac{1}{1+4d\eps}\br{\sum_{c'} \, \bigg| \, \frac{4d\eps{-}\alpha_{c'}}{d} \,
 \biggl( \sum_cv_c \biggr) \, \bigg|} + \frac{4d\eps}{1+4d\eps}\\
&\leq& \sum_{c'}\bigg|\frac{4d\eps- \alpha_{c'}}{d}\bigg|+4d\eps \\
&\overset{(b)}\leq& 8d\eps + 4d\eps \leq 12d\eps\\
\enqarst
In the above, step $(a)$ uses the fact that $vM$ is a probability distribution so $\|vM\|_1=1$ and step $(b)$ uses (\ref{eq:errorst1}) to substitute $\sum_{c'}|\alpha_{c'}|\leq 4d\eps$. This completes the proof.
\end{proof}

Now, we are ready to prove our main result, using the condition 
$\|v-vM\|_1 \leq 4\eps$ 
from (\ref{eq:errorst}). 
\begin{theorem}
\label{thm:approxcase}
Let $d \geq 2$ \change{and $M$ as defined before.}
It holds that 
$$\forall_c ~~M_{c,c}\geq 1-24d^4\eps.$$
\end{theorem}
\begin{proof}
Let $N$ be the doubly stochastic matrix as promised in Lemma \ref{lem:appsto}. We start with
\beqar
\label{eq:Ndeltaclose}
\|v- vN\|_1&\leq& \|v-vM\|_1 + \|vM-vN\|_1\nonumber\\
&\overset{(a)}\leq& 4\eps + 12d\eps \leq 16d\eps.
\enqar
Here, $(a)$ uses (\ref{eq:errorst}) and Lemma \ref{lem:appsto}.

Now, we apply the Birkoff-von Neumann theorem to write $N=\sum_{i=1}^k
q_i \Pi_i$, where $\Pi_1=\id$ and $\Pi_2, \cdots, \Pi_k$ are
permutation matrices not equal to the identity.  
If $q_1 = 1$ then we are done
so assume $q_1 < 1$, and there exists $i \geq 2$ with $q_i > 0$.
We lower bound $\|v-vN\|_1$ as follows.
\beqar
\label{eq:lb1}
\|v-vN\|_1&=&\bigg\|v-\sum_{i=1}^kq_i \br{v\Pi_i}\bigg\|_1 = \bigg\|v-q_1v - \sum_{i=2}^k q_i\br{v\Pi_i}\bigg\|_1 \nonumber\\
&=& \bigg\|(1-q_1)v - \sum_{i=2}^k q_i\br{v\Pi_i}\bigg\|_1 = (1-q_1)\bigg\|v - \sum_{i=2}^k \frac{q_i}{1-q_1}\br{v\Pi_i}\bigg\|_1.
\enqar 
The last expression in  (\ref{eq:lb1}) is lower bounded by the minimum $\ell_1$ distance between the vector $v$ and the convex hull of the vectors $\{v\Pi_i\}_{i=2}^k$. As shown in Claim \ref{clm:linprog} below, this distance is lower bounded by
$$\bigg\|v - \sum_{i=2}^k \frac{q_i}{1-q_1}\br{v\Pi_i}\bigg\|_1\geq \frac{1}{v_1}\br{\sum_jv^2_j-\max_{i>1}\br{\sum_jv_j\br{v\Pi_i}_j}}.$$ From Claim \ref{clm:permoverlap}, 
$$\max_{i>1}\br{\sum_jv_j\br{v\Pi_i}_j} \leq \sum_iv_i^2 - \frac{1}{\eta^2}.$$ This implies
$$\bigg\|v - \sum_{i=2}^k \frac{q_i}{1-q_1}\br{v\Pi_i}\bigg\|_1\geq \frac{1}{v_1\eta^2}.$$ From  (\ref{eq:lb1}), we conclude
$$\|v-vN\|_1\geq (1-q_1)\frac{1}{v_1\eta^2},$$
and combining this with  (\ref{eq:Ndeltaclose}), we find
\beq
\label{eq:lowprob}
(1-q_1)\frac{1}{v_1\eta^2} \leq 16d\eps \implies 1-q_1 \leq 16d\eps v_1\eta^2 \leq 16d^4\eps.
\enq
Since $N_{c,c}\geq q_1$, we finally conclude from Lemma \ref{lem:appsto} and  (\ref{eq:lowprob}) that
$$M_{c,c}\geq N_{c,c} - 12d\eps \geq q_1 - 12d\eps\geq 1- 24d^4\eps.$$
Here we used $d\geq 2$ to substitute $16d^4\eps + 12d\eps \leq 24d^4\eps$. This completes the proof.
\end{proof}

The following claims were used in the proof.
\begin{claim}
\label{clm:linprog}
Let $v, w_1, \ldots w_s$ be vectors in $\bR^n$ with non-negative entries. Suppose $v_1$ is the largest entry of $v$, and $v_1 > 0$. It holds that
$$\min_{r_1, \ldots r_s \in [0,1]: \sum_i r_i=1}\|v-\sum_ir_iw_i\|_1 \geq \frac{1}{v_1}\br{\sum_jv^2_j-\max_i\br{\sum_jv_j\br{w_i}_j}}$$
\end{claim}
\begin{proof}
We use the fact that $\|u\|_1= \max_{m_1,\ldots m_n\in [-1,1]} \sum_j u_jm_j$ to write
\beqarst
&&\min_{r_1, \ldots r_s \in [0,1]: \sum_i r_i=1}\|v-\sum_ir_iw_i\|_1\\
&&=\min_{r_1, \ldots r_s \in [0,1]: \sum_i r_i=1}\quad\max_{m_1,\ldots m_n\in [-1,1]}\br{\sum_jm_j\br{v_j-\sum_{i}r_i\br{w_i}_j}}\\
&&=\min_{r_1, \ldots r_s \in [0,1]: \sum_i r_i=1}\quad\max_{m_1,\ldots m_n\in [-1,1]}\br{\sum_jm_jv_j-\sum_{i,j}r_im_j\br{w_i}_j}\\
&&\overset{(a)}=\max_{m_1,\ldots m_n\in [-1,1]}\quad\min_{r_1, \ldots r_s \in [0,1]: \sum_i r_i=1}\br{\sum_jm_jv_j-\sum_{i,j}r_im_j\br{w_i}_j}\\
&&\overset{(b)}= \max_{m_1,\ldots m_n\in [-1,1]}\br{\sum_jm_jv_j-\max_i\br{\sum_jm_j\br{w_i}_j}}\\
&& \overset{(c)}\geq \frac{1}{v_1}\br{\sum_jv^2_j-\max_i\br{\sum_jv_j\br{w_i}_j}}.
\enqarst
Here $(a)$ uses the minimax theorem \cite{vonNeumann1928}, $(b)$ chooses the optimal $\{r_1, \ldots r_s\}$ to maximize $$\sum_{i,j}r_im_j\br{w_i}_j = \sum_{i}r_i\br{\sum_jm_j\br{w_i}_j}$$ and $(c)$ chooses $m_i=\frac{v_i}{v_1} \in [0,1]$. This completes the proof.
\end{proof}

\begin{claim}
\label{clm:permoverlap}
\change{Let $\{v_1, v_2, \ldots v_n\}$ be a vector and define $\gamma:=\min_{i\neq j}|v_i-v_j|$.} It holds that
$$\max_{\Pi\neq \id}\br{\sum_j v_j\br{v\Pi}_j}= \sum_jv_j^2 -  \gamma^2 ,$$
where the maximization is over all permutations $\Pi$ not equal to identity. \change{In particular, for the vector $v$ in our ensemble (see Equation \ref{eq:ourdist}, with $n$ replaced by $d$), we have $\gamma=\frac{1}{\eta}$ and $$\max_{\Pi\neq \id}\br{\sum_j v_j\br{v\Pi}_j}= \sum_jv_j^2 -  \frac{1}{\eta^2}.$$}
\end{claim}
\begin{proof}
We recall that $\Pi$ can be decomposed as a product of disjoint cycles $C_1, C_2, \ldots C_s$. If $\Pi$ leaves an index unchanged, we will view the action of $\Pi$ on this index as a `trivial cycle' of size $1$. Let $L_k$ be the set of indices on which the cycle $C_k$ acts. Since the cycles are disjoint,
\beq
\label{eq:cycperm}
\sum_j v_j\br{v\Pi}_j = \sum_{k=1}^s \sum_{j\in L_k}v_j\br{vC_k}_j.
\enq
Consider the expression $\sum_{j\in L_k}v_j\br{vC_k}_j$. Let $|L_k|$ be the number of elements in $L_k$. Relabel the elements in $L_k$ with integers $\{0,1,\ldots |L_k|-1\}$ in a manner that $C_k(m)=m+s \mod |L_k|$, for some integer $s$. Then  
\beqarst
\sum_{j\in L_k}v_j\br{vC_k}_j &=& \sum_{m=0}^{|L_k|-1} v_mv_{m+s \mod |L_k|} \\
&=& \frac{1}{2}\sum_{m=0}^{|L_k|-1} \br{v^2_m+v^2_{m+s \mod |L_k|} - \br{v_m-v_{m+s \mod |L_k|}}^2}\\
&=& \sum_{m=0}^{|L_k|-1}v_m^2 - \frac{1}{2}\sum_{m=0}^{|L_k|-1}(v_m-v_{m+s \mod |L_k|})^2\\
&\overset{(a)}\leq& \sum_{m=0}^{|L_k|-1}v_m^2 - \gamma^2\cdot\lfloor|L_k|/2\rfloor = \sum_{j\in L_k}v_j^2 - \gamma^2\cdot\lfloor |L_k|/2\rfloor.
\enqarst
Here $(a)$ follows by noticing that $(v_m-v_{m+s \mod |L_k|})^2\geq \gamma^2$. Combining with  (\ref{eq:cycperm}), we find that
$$\sum_j v_j\br{v\Pi}_j \leq \sum_{k=1}^s\br{\sum_{j\in L_k}v_j^2 -  \gamma^2\cdot\lfloor|L_k|/2\rfloor} = \sum_jv_j^2 -  \gamma^2\cdot\br{\sum_{k=1}^s\lfloor|L_k|/2\rfloor}.$$
Since $\Pi$ is not the identity permutation, there is a $C_k$ of length at least $2$. Thus, $\sum_{k=1}^s\lfloor|L_k|/2\rfloor\geq 1$ and
$$\max_{\Pi}\br{\sum_j v_j\br{v\Pi}_j}\leq \sum_jv_j^2 -  \gamma^2.$$
To show that the right hand side is achieved, choose $\Pi$ to be any permutation that swaps a pair of consecutive indices and leaves every other index unchanged. This completes the proof.
\end{proof}

\subsection{Final lower bound}

Theorem \ref{thm:approxcase} says that $M_{c,c} \geq 1-24d^4\eps$ for
all $c$. If we choose $\eps= \frac{1}{24 d^4\log d}$, we conclude that 
for all $c$
$$M_{c,c} \geq 1-\frac{1}{\log d}.$$ 
Thus, the lower bound in Claim \ref{clm:genlowb} takes the form
$$\condmutinf{C}{C'}{X}_\tau \; \geq \; \cent{C}{X}_{\rho} -2 -\br{1-\sum_cp_C(c)M_{c,c} + \eps }\log\br{d} \; \geq \; \cent{C}{X}_{\rho}-4.$$
The conditional entropy can be evaluated to be
\begin{eqnarray*}
\cent{C}{X}_{\rho} &=& \frac{1}{2}\br{\ent{C}_{\rho^0} + \ent{C}_{\rho^1}}\\
&=& \frac{1}{2}\log d +\frac{1}{2}\br{\sum_c v_c \log\frac{1}{v_c}}\\
 &=& \frac{1}{2}\log d +\frac{1}{2}\br{\sum_c v_c \log\frac{\eta}{d-c+1}}\\
&=& \frac{1}{2}\log d +\frac{1}{2}\br{\log\eta- \sum_c v_c \log(d-c+1)}\\
&\overset{(a)}\geq& \frac{1}{2}\log d +\frac{1}{2}\br{\log\eta- \log\br{\sum_c v_c (d-c+1)}}\\
&=& \frac{1}{2}\log d +\frac{1}{2}\br{\log\eta- \log\br{\sum_c \frac{(d-c+1)^2}{\eta}}}\\
&\overset{(b)}=& \frac{1}{2}\log d +\frac{1}{2}\br{\log\frac{d(d+1)}{2}- \log\br{\frac{d(d+1)(2d+1)}{3d(d+1)}}}\\
&=& \frac{1}{2}\log d +\frac{1}{2}\br{\log\frac{3d(d+1)}{4d+2}}\geq (\log d)-1.\\
\end{eqnarray*}
Here $(a)$ follows from concavity of log and $(b)$ follows by substituting the value of $\eta$. This leads to the lower bound
$$\condmutinf{C}{C'}{X}_\tau \geq (\log d)-5.$$
Thus, from Theorem \ref{thm:mixedlb}, we have a state $\rho_{XC}$ such that the asymptotic rate of communication for local error $\eps= \frac{1}{24 d^4\log d}$ is at least $(\log d)-7 $ while the Holevo information is at most $1$. 
Note that $d$ is independent of the number of copies $n$, and $\eps$ is independent of $n$ and only depends on $d$.  

\subsection*{Acknowledgements}
We thank Masato Koashi and Andreas Winter for helpful discussions. This work was completed when AA was affiliated to IQC and C\&O, University of Waterloo and the Perimeter Institute for Theoretical Physics, Waterloo.
This research is supported by research grants administered by NSERC and CIFAR.  
Perimeter Institute is supported in part by the Government of Canada and the Province of Ontario.

\bibliographystyle{ieeetr}
\bibliography{References}

\appendix

\section{Koashi-Imoto characterization at non-vanishing error}
\label{koashiIrobust}

Koashi and Imoto \cite{KoashiI1, KoashiI2, KoashiI3} provide a 
characterization of the optimal rates of quantum communication, 
classical communication, and entanglement for 
$(n, R, \eps)$- blind compression of general quantum states
(defined similarly to the task of blind distribution compression) in
the limit where $n \rightarrow \infty$ and the local error 
$\eps \rightarrow 0$.  
These results apply to both the unassisted and the
entanglement-assisted scenarios.

The Koashi-Imoto characterization of $(n, R, \eps)$- blind compression
is achieved by the following structure theorem for any ensemble
$\{p_i, \rho_i\}$. There exists a decomposition of the Hilbert space
$\cH$ as $\cH= \bigoplus_{\ell=1}^L \cH^{(\ell)}_J\otimes \cH^{(\ell)}_K$,
in a manner that $\rho_i=\bigoplus_{\ell=1}^L q_{i,\ell}
\rho^{(i,\ell)}_J\otimes \rho^{\ell}_K$.  
Here, $\rho^{(i,\ell)}_J$ and $\rho^{\ell}_K$ are normalized density
matrices acting on $\cH^{(\ell)}_J$ and $\cH^{(\ell)}_K$ respectively, 
and $q_{i,\ell} \geq 0$, $\sum_{\ell=1}^L q_{i,\ell} = 1$.  Furthermore, 
for each $\ell$, $\{\rho^{(1,\ell)}_J, \rho^{(2,\ell)}_J, \cdots \}$
cannot be expressed in a simultaneously block-diagonal form.
The register $K$ is viewed
as redundant, as it has no dependence on $i$.  Without loss of
generality, one can restrict the attention to ensembles which have no
redundant register $K$, as the register $K$ can be removed by Alice
and later be generated by Bob without communication between them.
The average state $\rho = \sum_i p_i \rho_i$ can also be written as
$\rho =\bigoplus_{\ell=1}^L p^{(\ell)} \rho^{(\ell)}_J$ where $p^{(\ell)}
= \sum_i p_i q_{i,\ell}$ and $\rho^{(\ell)}_J = {1 \over p^{(\ell)}}
\br{\sum_i p_i q_{i,\ell} \rho^{(i,\ell)}_J}$.  Furthermore, the authors proved that
any channel $\cE$ satisfying $\cE(\rho_i)=\rho_i$ acts as
the identity map on $J$. 

The characterization in the case of zero error ($\eps=0$) is now immediate. Let $\cP$ be the combined encoding and decoding map in the protocol. Since the protocol makes no error, $\cP(\rho_i)=\rho_i$ for all $i$ and hence $\cP$ acts as the identity map on the $J$ register, \change{leading to $\log (|J|L)$ as the optimal rate in this case}.

%
To address the regime $n\rightarrow \infty, \eps\rightarrow 0$, 
Koashi and Imoto \cite{KoashiI2} \change{define $I_C = - \sum_\ell p^{(\ell)} \log p^{(\ell)}$ and 
$I_Q = \sum_\ell p^{(\ell)} S\br{\rho^{(\ell)}_J}$.  
The ensemble can be compressed at the rates of $I_C$ bits and 
$I_Q$ qubits combined, whereas $I_C+I_Q = S(\rho)$. To prove optimality of this rate, they} introduce two error functions
$$f(\cP) = 1- \sum_ip_i\F(\rho_i, \cP(\rho_i))$$ and $$g(\cP)=H(\lambda) + \lambda\log (d-1),$$ where $\lambda=1- \sum_a r(a)\Tr(\ketbra{a} \, \cP(\ketbra{a}))$ and $\ket{a}, r(a)$ are the eigenvectors and eigenvalues of the average state of the ensemble, $\rho=\sum_ip_i\rho_i$. The parameter $f(\cP)$ captures the error $\eps$ (up to the difference that we are working in trace distance, whereas they work in fidelity). The parameter $g(\cP)$ appears in their lower bound $R\geq S(\rho) - g(\cP)$ on the communication cost. They give a continuity argument that if $f(\cP)\rightarrow 0$, then $g(\cP)\rightarrow 0$. 

For non-vanishing error, the dependence of $\lambda$ on $f(\cP)$ and
in turn, the dependence of $g(\cP)$ on $f(\cP)$ (via $\lambda$) is
important.  While $\lambda \rightarrow 0$ as $f(\cP) \rightarrow 0$,
it is unclear how quickly $\lambda$ vanishes as $f(\cP) \rightarrow
0$.  Consequently, $g(\cP)$ need not vanish quickly enough to
provide a strong lower bound.
In our context, we construct a protocol in Theorem
\ref{theo:lowcommprot} exhibiting this sensitivity. We show an example
where $g(\cP)$ becomes close to $\log d$, even if $f(\cP) \approx
\frac{1}{\sqrt{d}}$.
Thus we know from the example that for two distributions, the lower
bound proved by Koashi and Imoto only works for error substantially
smaller than ${1 \over \sqrt{d}}$.

Another reason to expect that the Koashi-Imoto characterization is sensitive to errors, is as follows.  Consider an equiprobable ensemble of two states $\rho$ and $\sigma$.  When the error $\eps$ (defined in (\ref{eq:locerror})) is finite, we have the conditions $\Delta(\cP(\rho), \rho) \leq 2\eps, \Delta(\cP(\sigma), \sigma) \leq 2\eps$. Does this guarantee that $\rho$ and $\sigma$ are respectively close to states $\tilde{\rho}$ and $\tilde{\sigma}$ that are fixed points of $\cP$?  If this were the case, we could apply the Koashi-Imoto decomposition to $\tilde{\rho}, \tilde{\sigma}$ to obtain a large lower bound for the rate for finite $\eps$. Unfortunately this is not true, as witnessed by the channel $\cE(\tau)= (1-\eps)\tau + \eps \frac{\id}{d}$. This channel satisfies $\Delta(\cE(\tau), \tau)\leq \eps$ for all $\tau$, but its only fixed point is $\frac{\id}{d}$. 

Our arguments are made precise in the following protocol for the blind compression of any pair of commuting states $\rho,\sigma$, with rate significantly less than $\log(d)$ for non vanishing local error. 
\begin{theorem}
\label{theo:lowcommprot}
Let $n$ be a positive integer and $\delta\in (0, \frac{1}{2}),\gamma \in (0,1)$. Consider the local error model.  Given two commuting states $\rho, \sigma$, there exists an $(n,R,\delta+ \gamma)$ blind distribution compression protocol $\cP$ with $R = 2\log\log(\frac{d}{\gamma}) + 2\log\frac{1}{\delta}+3$.
\end{theorem}
\begin{proof}

Without loss of generality, $\rho, \sigma$ are both diagonal in the
computational basis, so, they can be written as $\rho=\sum_a p(a) \ketbra{a}$ and $\sigma=\sum_a q(a) \ketbra{a}$ where $p(a), q(a) \geq 0$, $\sum_a p(a) = \sum_a q(a) = 1$. Let 
$u = \Bigl \lceil \frac{\log\frac{d}{\gamma}}{\log\frac{1}{1-\delta}} \Bigr \rceil$, so that $(1-\delta)^u \leq \frac{\gamma}{d}$.  
For $i,j \in \{1,2,\cdots,u\}$, define the sets 
\begin{align}
T_{i,j} &:= \{ \, a \, : \, p(a) \! \in \! ((1{-}\delta)^{i}, (1{-}\delta)^{i-1}] \, , \; q(a) \! \in \! ((1{-}\delta)^{j}, (1{-}\delta)^{j-1}] \, \} \, ,
\nonumber \\ 
T_{i,u+1} &:= \{ \, a \, : \, p(a) \! \in \! ((1{-}\delta)^{i}, (1{-}\delta)^{i-1}]\, , \; q(a) \leq (1{-}\delta)^{u} \, \} \,,
\nonumber \\
T_{u+1,j} &:= \{ \, a: \, p(a) \leq (1{-}\delta)^{u} \,, \; q(a) \! \in \! ((1{-}\delta)^{j}, (1{-}\delta)^{j-1}]\}
\nonumber \\
T_{u+1,u+1} &:= \{ \, a: \, p(a) \leq (1{-}\delta)^{u} \,, \; q(a) \leq (1{-}\delta)^{u}\} \,.
\nonumber
\end{align}
The protocol $\cP$ is as follows. 
\begin{itemize}
\item Alice receives a sample $a$. She finds the unique $(i,j)$ such that $a\in T_{i,j}$ and communicates $(i,j)$ to Bob.
\item Receiving $(i,j)$, Bob outputs an $a'$ drawn uniformly from $T_{i,j}$.  
\end{itemize}
If the input given to Alice is drawn from $\rho$ ($\sigma$), let the output produced by Bob be drawn from $\rho'$ ($\sigma'$).  
The analysis of the protocol is as follows.
\begin{itemize}
\item {\bf Communication cost}: Since there are at most $(u+1)^2$ $(i,j)$'s, it suffices for Alice to communicate
$2\log(u+1) \leq 2 ((\log u) + 1) \leq 2 \log \log
  \frac{d}{\gamma} + 2 \log \frac{1}{\log\frac{1}{1{-}\delta}} + 3
  \leq 2\log\log\frac{d}{\gamma} + 2\log\frac{1}{\delta} + 3$
bits to Bob.  
Note that we have used the inequality $\frac{1}{-\log(1-\delta)} \leq \frac{1}{\delta}$ derived from the Taylor series expansion of $\log(1-\delta)$.   
\item {\bf Error analysis}: Let $p(T_{i,j}) := \sum_{a\in T_{i,j}}p(a)$ and $q(T_{i,j}) := \sum_{a\in T_{i,j}}q(a)$. 
We can rewrite $\rho$ and $\sigma$ as
$$\rho= \sum_{i,j=1}^{u+1}p(T_{i,j}) \sum_{a\in T_{i,j}}\frac{p(a)}{p(T_{i,j})}\ketbra{a}, \quad \sigma= \sum_{i,j=1}^{u+1}q(T_{i,j}) \sum_{a\in T_{i,j}}\frac{q(a)}{q(T_{i,j})}\ketbra{a}.$$ 
Observe from the protocol that
$$\rho'= \sum_{i,j=1}^{u+1}p(T_{i,j}) \sum_{a\in T_{i,j}}\frac{1}{|T_{i,j}|}\ketbra{a}, \quad \sigma' = \sum_{i,j=1}^{u+1}q(T_{i,j}) \sum_{a\in T_{i,j}}\frac{1}{|T_{i,j}|}\ketbra{a}.$$ 
So, 
\begin{eqnarray}
\Delta(\rho, \rho')&=&\Delta\bigg(\sum_{i,j=1}^{u+1}p(T_{i,j})\sum_{a\in T_{i,j}}\frac{p(a)}{p(T_{i,j})}\ketbra{a} \, , \; \sum_{i,j=1}^{u+1}p(T_{i,j})\sum_{a\in T_{i,j}}\frac{1}{|T_{i,j}|}\ketbra{a}\bigg)\nonumber\\
&=& \sum_{i,j=1}^{u+1} p(T_{i,j}) \; \Delta\bigg(\sum_{a\in T_{i,j}}\frac{p(a)}{p(T_{i,j})}\ketbra{a} \, , \; \sum_{a\in T_{i,j}}\frac{1}{|T_{i,j}|}\ketbra{a}\bigg)
\label{unifinset}
\end{eqnarray}
By definition of the set $T_{i,j}$ with $i\leq u$, it holds that $1-\delta\leq\frac{p(a_1)}{p(a_2)}\leq \frac{1}{1-\delta}$ for all $a_1, a_2 \in T_{i,j}$. Thus, for all $a\in T_{i,j}$ with $i\leq u$,
\begin{equation}
(1-\delta)\frac{1}{|T_{i,j}|}\leq \frac{p(a)}{p(T_{i,j})}\leq\frac{1}{1-\delta}\frac{1}{|T_{i,j}|}\implies \bigg|\frac{p(a)}{p(T_{i,j})}- \frac{1}{|T_{i,j}|}\bigg| \leq \frac{\delta}{1-\delta}\frac{1}{|T_{i,j}|} \,.
\nonumber
\end{equation}
So, 
\begin{equation}
\Delta\bigg(\sum_{a\in T_{i,j}}\frac{p(a)}{p(T_{i,j})}\ketbra{a} \, , \, \sum_{a\in T_{i,j}}\frac{1}{|T_{i,j}|}\ketbra{a}\bigg) 
\leq {1 \over 2} \sum_{a \in T_{i,j}} \frac{\delta}{1-\delta}\frac{1}{|T_{i,j}|} \leq \frac{\delta}{2(1-\delta)} \,.
\label{eq:closeratios}
\end{equation}
Furthermore, 
\begin{equation}
\sum_{j=1}^{u+1}p(T_{u+1,j})= \sum_{a: p(a) \leq (1-\delta)^u}p(a)\leq (1-\delta)^u\cdot d \leq \gamma \,.
\label{eq:truncation}
\end{equation}
Applying (\ref{eq:closeratios}), (\ref{eq:truncation}), and $\delta \leq {1 \over 2}$ to (\ref{unifinset}), 
\begin{eqnarray}
\Delta(\rho, \rho') 
\; \leq \; \sum_{i=1}^u\sum_{j=1}^{u+1} \, p(T_{i,j}) \, \frac{\delta}{2(1-\delta)} +  \sum_{j=1}^{u+1}p(T_{u+1,j})
\; \leq \;  \frac{\delta}{2(1-\delta)} + \gamma 
\; \leq \; \delta + \gamma \,.
\nonumber 
\end{eqnarray}
A similar argument shows that $\Delta(\sigma,\sigma') \leq \delta+\gamma.$ This completes the error analysis.
\end{itemize}
The correctness of the protocol concludes by running the above protocol independently for each copy to obtain an $(n,R, \delta+\gamma)$ protocol. 
\end{proof}

Theorem \ref{theo:lowcommprot} immediately implies that $g(\cP)$ is close to $S(\rho)$ for constant $\delta, \gamma$, else the lower bound $R\geq S(\rho)- g(\cP)$ would contradict the statement of the theorem. We can see this explicitly by evaluating the functions $f(\cP)$ and $g(\cP)$. Let us continue using the notation in Theorem \ref{theo:lowcommprot}.  Since $\Delta(\rho, \cP(\rho))\leq \delta+\gamma$, the Fuchs-van de graaf inequality implies that $\F(\rho, \cP(\rho))\geq 1-2\delta-2\gamma$. Similarly, $\F(\sigma, \cP(\sigma))\geq 1-2\delta-2\gamma$. Hence $f(\cP)\leq 2\delta+2\gamma$. Now, for the given ensemble, $r(a)=\frac{1}{2}p(a)+\frac{1}{2}q(a)$. 
Then, 
\begin{eqnarray*}
1-\lambda&=& \sum_a r(a) \, \Tr(\ketbra{a}\cdot\cP(\ketbra{a}))\\
&=& \sum_{i,j=1}^{u+1}\sum_{a\in T_{i,j}} r(a) \, \Tr\br{\ketbra{a}\cdot \br{\sum_{a'\in T_{i,j}}\frac{\ketbra{a'}}{|T_{i,j}|}}}\\
&=& \sum_{i,j=1}^{u+1}\sum_{a\in T_{i,j}} r(a)\cdot\frac{1}{|T_{i,j}|}
\leq \max_{a'} \, r(a') \cdot \sum_{i,j=1}^{u+1}\sum_{a\in T_{i,j}}\frac{1}{|T_{i,j}|}\\
&=& \max_{a'} \, r(a')\cdot (u+1)^2 \leq \frac{4\log^2\frac{d}{\gamma}}{\delta^2}\cdot\max_{a'} \, r(a'). 
\end{eqnarray*}
Suppose $p$ and $q$ are the uniform and the staircase distributions (Figure \ref{fig:exampledist}). That is, $p(a)=\frac{1}{d}$ for all $a$ and $q(a)= \frac{2(d-a+1)}{d(d+1)}=\frac{2}{d}-\frac{2a}{d(d+1)}$ (note that these distributions have no redundant part). Then $\max_a r(a) \leq \frac{3}{2d}$, which leads to $1- \lambda \leq \frac{6\log^2\frac{d}{\gamma}}{d\delta^2}$. This gives us $g(\cP) \geq \lambda \log(d{-}1) \geq \log(d{-}1) - \frac{6\log^3\frac{d}{\gamma}}{d\delta^2}$. If $\delta=\gamma = \frac{\log^2d}{\sqrt{d}},$ then $g(\cP)\geq \log(d{-}1) - \frac{20}{\log(d)}$ and $f(\cP) \leq \frac{4\log^2d}{\sqrt{d}}$.  So, $g(\cP)$ is close to $\log d$, even when $f(\cP) \approx \frac{1}{\sqrt{d}}$.

\end{document}